\documentclass[aps,prl,showpacs,superscriptaddress,twocolumn] {revtex4}
\usepackage{graphicx}

\begin{document}

\title{Observation of a continuous phase transition in a shape-memory alloy}

\author {J. C. Lashley}
\affiliation{Los Alamos National Laboratory, Los Alamos, NM 87545,
USA}

\author{S. M. Shapiro}
\affiliation{Brookhaven National Laboratory, Upton, NY 11973, USA}

\author{B. L. Winn}
\affiliation{Oak Ridge National Laboratory, Oak Ridge, TN 37831, USA}

\author{C. P. Opeil}
\affiliation{Boston College, Department of Physics, 140 Commonwealth
Avenue, Chestnut Hill, MA 02467, USA}

\author{M. E. Manley}
\affiliation{Lawrence Livermore National Laboratory, Livermore, CA 94550, USA}

\author{A. Alatas}
\affiliation{Advanced Photon Source, Argonne National Laboratory,
Argonne, IL 60439, USA}

\author{W. Ratcliff}
\affiliation{National Institute of Standards and Technology,
Gaithersburg, MD 20899, USA}

\author{T. Park}
\affiliation{Los Alamos National Laboratory, Los Alamos, NM 87545,
USA} \affiliation{Department of Physics, Sungkyunkwan University,
Suwon 440-746, Korea}
\author{R. A. Fisher}
\affiliation{Los Alamos National Laboratory, Los Alamos, NM 87545,
USA}

\author{B. Mihaila}
\affiliation{Los Alamos National Laboratory, Los Alamos, NM 87545,
USA}
\author{P. Riseborough}
\affiliation{Department of Physics, Temple University, Philadelphia,
PA 19122, USA}

\author{E. K. H. Salje}
\affiliation{Department of Earth Sciences, University of Cambridge,
Downing Street, Cambridge CB2 3EQ, UK}

\author {J. L. Smith}
\affiliation{Los Alamos National Laboratory, Los Alamos, NM 87545,
USA}


\begin{abstract}

Elastic neutron-scattering, inelastic x-ray scattering,
specific-heat, and pressure-dependent electrical transport
measurements have been made on single crystals of AuZn and
Au$_{0.52}$Zn$_{0.48}$ above and below their martensitic transition
temperatures ($T_{M}$~=~64~K and 45~K, respectively). In each
composition, elastic neutron scattering detects new commensurate
Bragg peaks (modulation) appearing at $Q = (1.33,0.67,0)$ at
temperatures corresponding to each sample's $T_{M}$. Although the
new Bragg peaks appear in a discontinuous manner in the
Au$_{0.52}$Zn$_{0.48}$ sample, they appear in a continuous manner in
AuZn. Surprising us, the temperature dependence of the AuZn Bragg
peak intensity and the specific-heat jump near the transition
temperature are in favorable accord with a mean-field approximation.
A Landau-theory-based fit to the pressure dependence of the
transition temperature suggests the presence of a critical endpoint
in the AuZn phase diagram located at $T_M^*$~=~2.7~K and
$p^*$~=~3.1~GPa, with a quantum saturation temperature
$\theta_s$~=~48.3$\pm$3.7~K.
\end{abstract}


\maketitle

A class of materials exhibiting martensitic (diffusionless) phase
transformations yields properties used in a range of technological
applications including implants to increase flow in restricted blood
vessels~\cite{Duerig}, actuators for the treatment of high
myopia~\cite{Yu}, voltage generators~\cite{Suorsa}, and orthodontic
arch-wires~\cite{Jafari}. These properties often depend on the
history of the material and may allow it to recover its previous
shape after deformation, known as the shape-memory effect (SME). It
has long been recognized that these transformations are all
thermodynamically first order (discontinuous)~\cite{Albers, Boyd,
Gash}. Special cases arise when the order parameter is coupled to an
external field in a complicated way, leading to a weakly first-order
transition~\cite{Rubini}, which is believed to result from a
complicated coupling between strain and order parameter
fluctuations. Because shape-memory alloys and other multiferroic
materials owe their functionality to complicated cross-field
responses between two (or more) pairs of conjugate thermodynamic
variables, a description of the free-energy landscape and the
ability to predict functionality become one and the same.

Phenomenological descriptions of the martensitic transformation
pathway, based on reciprocal space~\cite{Bowles} and real
space~\cite{David} geometries, have established rules to determine
relative twin and crystallographic orientations between the
austenite (high-temperature) and martensite (low-temperature)
modifications. Knowledge of the symmetry breaking allows for a
definition of the energetic driving force in terms of the difference
in free energies between phases. In the Ginzburg-Landau (GL)
approach, the free energy, $G$, is expressed as a sum of symmetry
invariants. In its simplest form, $G$ is approximated by a
polynomial expansion in even powers of an order parameter, $Q$, as $
G( Q ) = aQ^2  + bQ^4 + cQ^6 + \frac{g}{2} | {\nabla Q } |^2$. For
martensitic transformations, $Q$ is taken to be strain or strain
coupled to shuffle (displacements involving quasi-static phonons
with fractional commensurate wave vectors with uniform
shears~\cite{Burgers}). The coefficients $a, b, c$, and $g$ are
material parameters to be determined experimentally or by first
principle simulations.

For the majority of shear-induced transformations, the GL expansion
captures the essential physics~\cite{Toledano,Salje}, such as
constitutive response~\cite{Avadh1,Iwata} and the occurrence of
anti-phase boundaries~\cite{Cao}. When applied to shape-memory
alloys, the presence of intervening (premartensitic)
phases~\cite{Sethna} presents difficulties for both experiment and
theory as to a unique characterization of the order parameter.
Further issues have arisen in materials exhibiting quantum
mechanical effects in the band structure and strong electron-phonon
coupling~\cite{Avadh2}. The thermodynamic behavior is best explored
when no further transitions occur between T$_M$ and 0~K. This effect
was recently reported for Fe-doped NiTi, where all thermodynamic
parameters could be correlated self-consistently within mean field
theory~\cite{Toni}. In order to integrate these cases into GL
theory, it is necessary to measure and access each thermodynamic
property and infer cross-field couplings of the order parameter. The
ability to distinguish subtle differences between first-order and
weakly first-order depends on the ability to resolve the behavior of
pertinent physical properties.

In this Letter we examine the AuZn system for thermodynamic
properties that contribute to the first-order or weakly first-order
nature of the free-energy landscape. We measure elastic
neutron-scattering, inelastic x-ray scattering, specific-heat, and
the pressure dependence of the transition by electrical transport.
While we observed classical first-order behavior at 45~K for
Au$_{0.52}$Zn$_{0.48}$, we observed a mean field continuous
transformation for AuZn near 64~K in the temperature dependence of
the Bragg peak intensity, and a Landau step with a $\lambda$ anomaly
in the excess specific heat.

\begin{figure}[!]
   \includegraphics[width=\columnwidth]{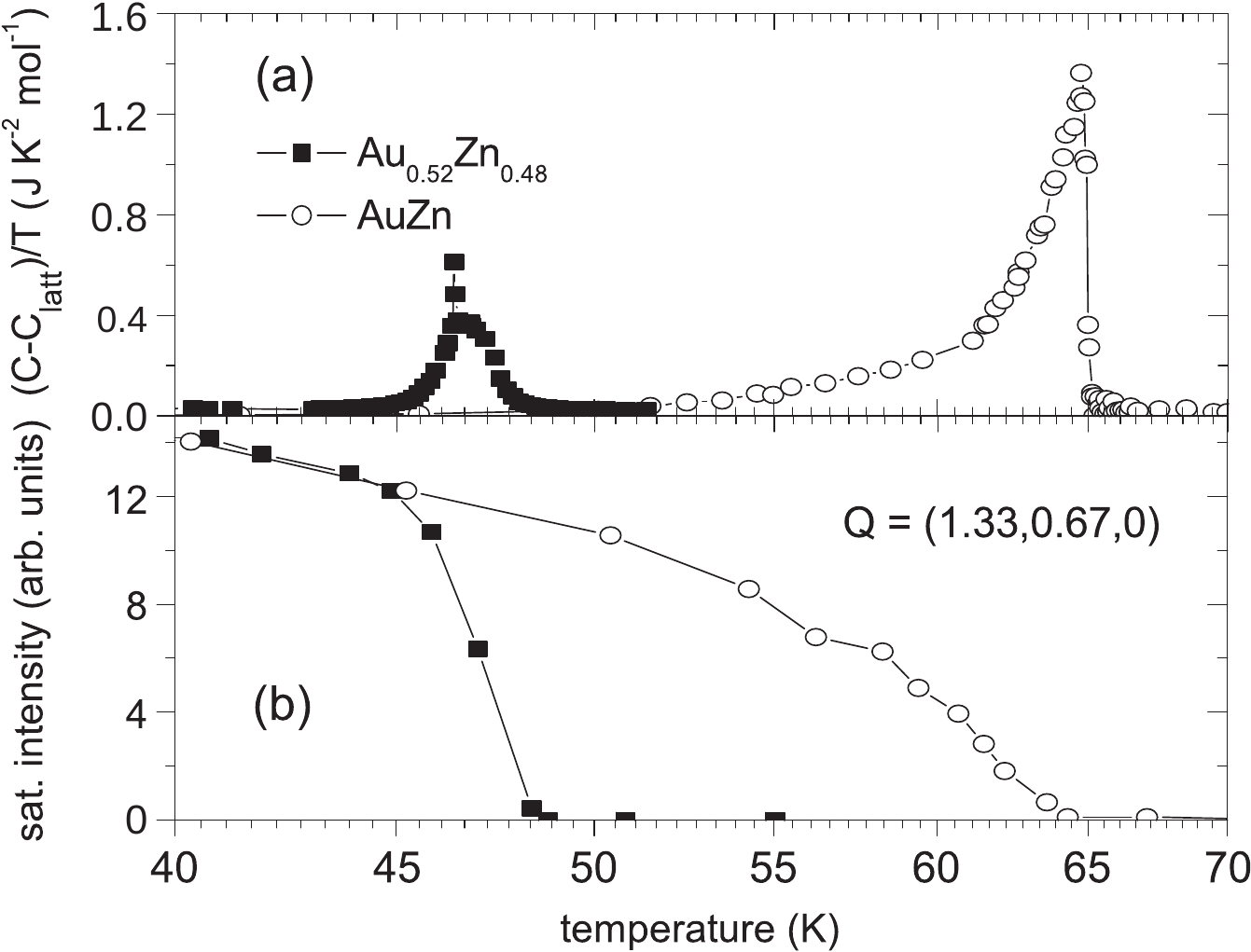}
   \caption{\label{Figure 1}
   (a) The excess specific heat divided by temperature versus temperature in the vicinity of
   the martensitic transitions for Au$_{0.52}$Zn$_{0.48}$ and AuZn.
   (b) Temperature dependence of the satellite-peak intensity along $Q = (1.33, 0.67, 0)$.
   The satellite peak intensity is proportional to the square of the order parameter.}
\end{figure}

\begin{figure}[b]
   \includegraphics[width=\columnwidth]{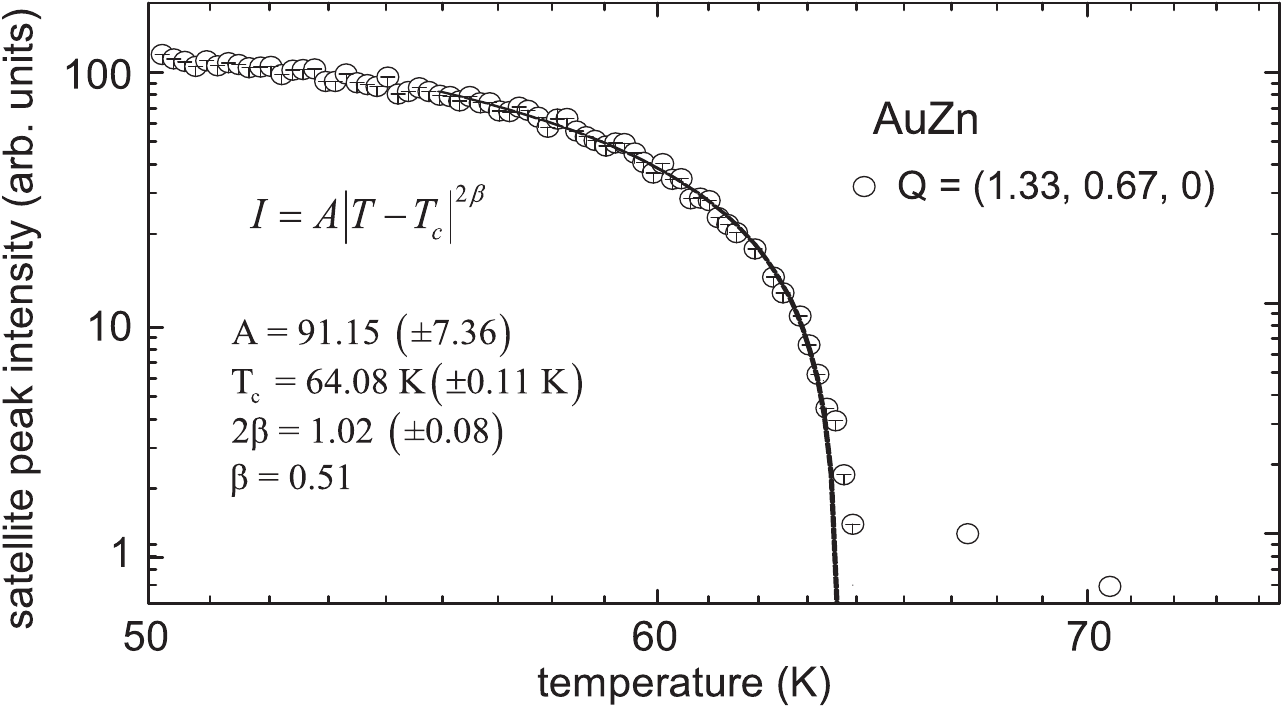}
   \caption{\label{Figure 2}
   Temperature dependence of the satellite peak intensity on heating along $Q = (1.33, 0.67, 0)$
   in AuZn, near the transition temperature. A fit to mean-field theory for $T \leq T_{c}$. The points above 64~K
are background.}
\end{figure}

Single crystals were prepared by fusion of the elements in a
Bridgman furnace and were oriented by back-reflection Laue
diffraction. The neutron experiments were performed on the BT9
triple-axis spectrometer at the NIST research reactor. Measurements
of the phonon dispersion were made on XOR of sector 3 at the
Advanced Photon Source, Argonne National
Laboratory~\cite{Sinn1,Sinn2}. Specific-heat measurements were
measured using a thermal-relaxation calorimeter by Quantum
Design~\cite{Lashley}. The pressure dependence of the transition
temperature was made by a four-terminal ac-transport method in a
mechanical pressure cell designed to reach pressures of 3~GPa.

\begin{figure}[!]
   \includegraphics[width=\columnwidth]{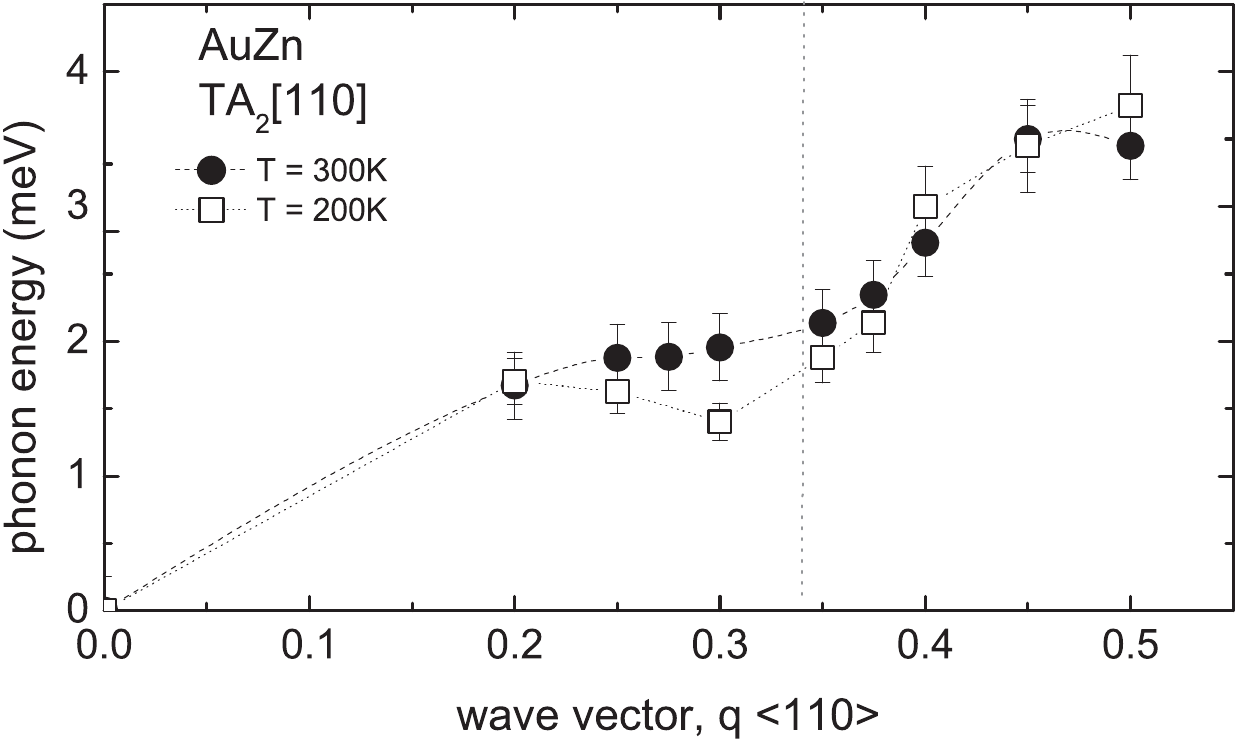}
   \caption{\label{Figure 3}
   Temperature dependence of the phonon dispersion in AuZn along the TA$_{2}$ branch.
   At 300~K one notices considerable phonon softening at $q = (0.33, 0.33, 0)$
   (vertical line). The dashed lines between points are provided as guides to the eye.}
\end{figure}

\begin{figure}[b]
   \includegraphics[width=\columnwidth]{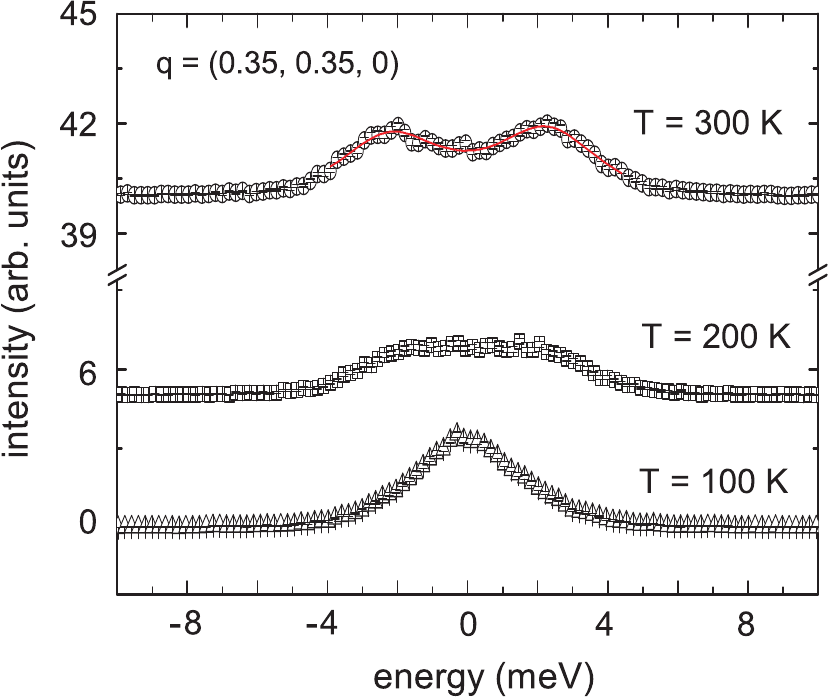}
   \caption{\label{Figure 4}
   Inelastic x-ray scattering data showing the temperature dependence near the soft mode, $q$~=~(0.35, 0.35, 0).
   The 300 and 200~K data are offset by 0.004 and 0.0005,
respectively on
   the y-axis for clarity. The energy positions are fit to a double Lorentzian (solid curve) at 300~K to determine the phonon energy.
   The intrinsic resolution for inelastic x-ray scattering is ~2 meV.}
\end{figure}

\begin{figure}[t]
   \includegraphics[width=\columnwidth]{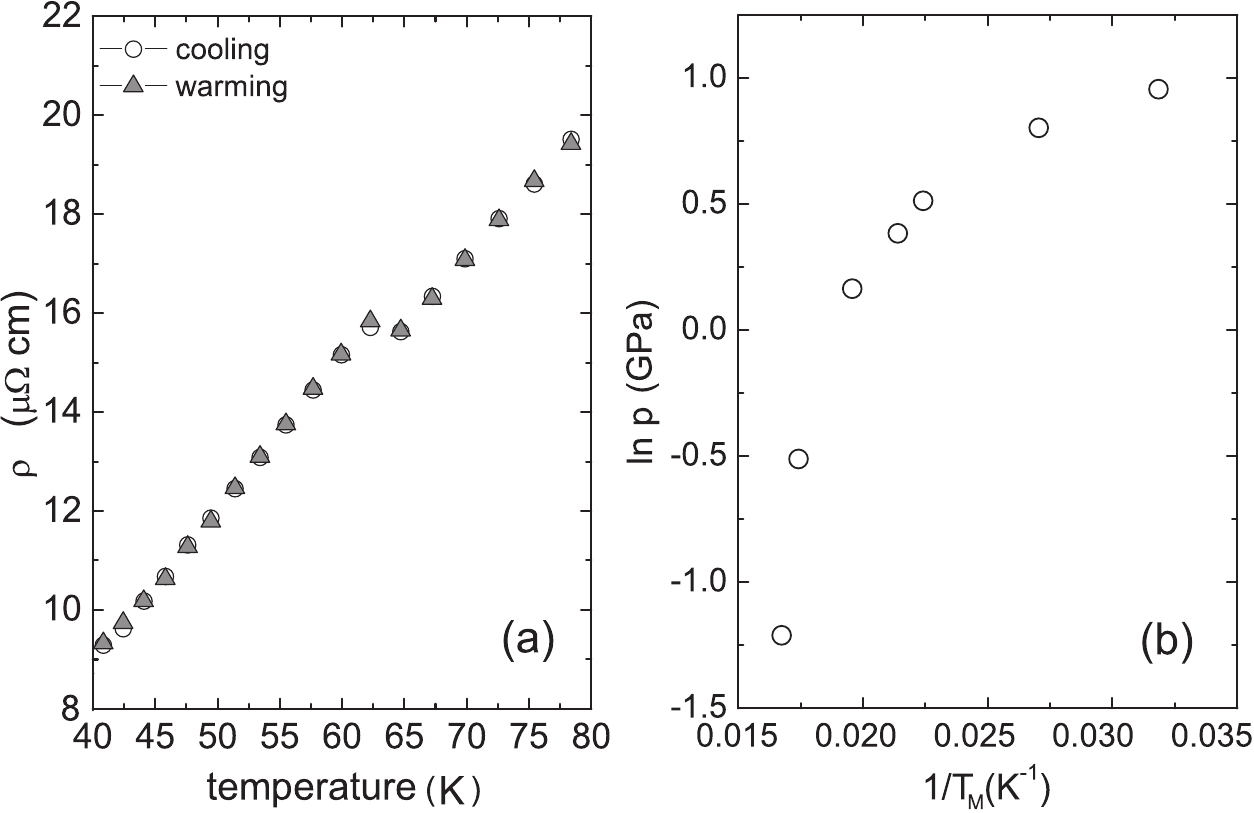}
   \caption{\label{Figure 5}
   (a) The electrical-resisitvity data taken on cooling and warming is shown in the vicinity of $T_M$. (b) Plot of $ln~p$ versus $1/T$ to test the applicability
Clausius-Clapeyron
   equation.
   }
\end{figure}

Previous investigations have shown that AuZn exhibits a martensitic
transition as seen with low-temperature electron
microscopy~\cite{Ted} and by recoverable transformation strain
(shape-memory effect)~\cite{Darling}. Figure~\ref{Figure 1} shows
the temperature dependence of (a) the excess specific heat and (b)
the satellite peak intensity for Au$_{0.52}$Zn$_{0.48}$ and AuZn.
Although we have measured the specific heat previously~\cite{Ross},
it is plotted along with the elastic neutron-scattering data, since
it is revealing to compare both sets of measurements that were made
on the same samples. We find a latent heat anomaly at 45~K for
Au$_{0.52}$Zn$_{0.48}$. A $\lambda$-anomaly was observed for AuZn,
but no latent heat was observed. The specific heat near the
transition can be influenced by intrinsic defects, including
triple-point and anti-structure defects, both known to be present in
the B2 structure~\cite{Austin, Gupta}. It was therefore necessary to
use a microscopic probe to determine whether or not the transition
could be a broadened first-order transition. Elastic
neutron-scattering measurements indicate that in each composition
new commensurate Bragg peaks (modulation) appear at $Q =
(1.33,0.67,0)$ at temperatures corresponding to each sample's
$T_{M}$~Figure~\ref{Figure 1}(b). Mirroring the specific-heat data,
the temperature dependence of the satellite peak intensity shows a
rapid jump at 45~K in Au$_{0.52}$Zn$_{0.48}$, while a continuous
variation occurs for AuZn on warming starting at 64~K. The satellite
peak intensity is proportional to the square of the order parameter,
since it leads to the low-temperature rhombohedral phase. The
specific-heat data also indicate thermal hysteresis (not shown) of
6~K at 45~K for Au$_{0.52}$Zn$_{0.48}$. The satellite peak intensity
for AuZn depicted in Fig.~\ref{Figure 1}(b) was measured by neutrons
on heating and cooling. Unlike the specific-heat data
Fig.~\ref{Figure 1}(a) we observe a small discontinuity and 1.5~K
difference in transition temperature on cooling.

Figure~\ref{Figure 2} shows the satellite peak intensity versus
temperature in the vicinity of $T_M$ for AuZn. Near the transition
temperature, the satellite peak evolves continuously with increasing
temperature, showing a temperature dependence given by $ I = A\left|
{T - T_M } \right|^{2\beta }; $ where the satellite peak intensity
is $I$, the critical temperature is $T_M$, and the critical exponent
is~$\beta$. We obtain T$_M$=64.08~K $\pm$0.11~K and
2$\beta$=~1.02$\pm$0.08. The value of the critical exponent obtained
from this fit ($\beta$=0.51) is close to the mean-field value of
$\beta$=0.50~\cite{Binney}. One could argue that mean-field
exponents were obtained by coincidence given that the critical
regime, as defined by the Ginzburg-criterion, is restricted to a
range of temperatures too small to be resolved by our experiments,
or that the correlation length does not diverge, since it is limited
by other extrinsic disorder (\textit{i.e.}, martensite twins) or
intrinsic fluctuations. The latter possibility seems more plausible
to us, because the inverse correlation length, or width of the Bragg
peaks (not shown), increases as the temperature is lowered below
$T_M$.

As the temperature is lowered, the unit cell is modulated in the
[110] shear direction. A commensurate shuffle of every third unit
cell results in a hexagonal primitive unit cell formed from nine
primitive cubic cells of the parent phase. This structure can also
be described in terms of its conventional rhombohedral unit cell. In
Fig.~\ref{Figure 3} we show the TA$_2$ phonon-dispersion curves
along [110], measured by inelastic x-ray scattering at temperatures
of 300~K and 200~K. The inflection in the phonon frequency near
$\xi=1/3$ indicates the low shear instability along the TA$_2$[110]
branch. This inflection is comparable to an earlier
investigation~\cite{Makita} of off-stoichiometric AuZn samples. At
300~K, the phonon energy positions have been fit over the energy
range -4~meV~$\leq E\leq$~4~meV to a double Lorentzian, as shown in
Figure~\ref{Figure 4}. The phonons continuously soften with
decreasing temperature to the point where separation becomes
difficult below 200~K.

Figure~\ref{Figure 5} (a) shows the change in resistivity between
cooling and warming. The data were collected with warming and
cooling rates of 0.2 K min$^{-1}$. In the region of $T_M$ there is
no hysteresis, to within experimental error. Because of the lack of
hysteresis, we elected to measure the pressure dependence of the
transition. For a first-order transition, the Clausius-Clapeyron
relation, $\ln p/p_0 = -\Delta H / (RT) + A$, predicts a linear
relationship for~$\ln p/p_0$ versus $1/T$ with the slope $\left( -
\Delta H / R \right)$. Here, $p$ denotes the pressure, $H$ the
enthalpy, $R$ the universal gas constant, and $p_0$ is a reference
pressure. Figure~\ref{Figure 5}(b) shows the plot of $\ln p$ versus
$1/T$. It appears that there is no observable linear region.
Assuming that the entropy does not vary significantly with pressure,
the conditions for the applicability of the Clausius-Clapeyron
relation are not met. This result provides further evidence that the
transition is not first order.

\begin{figure}[t]
   \includegraphics[width=\columnwidth]{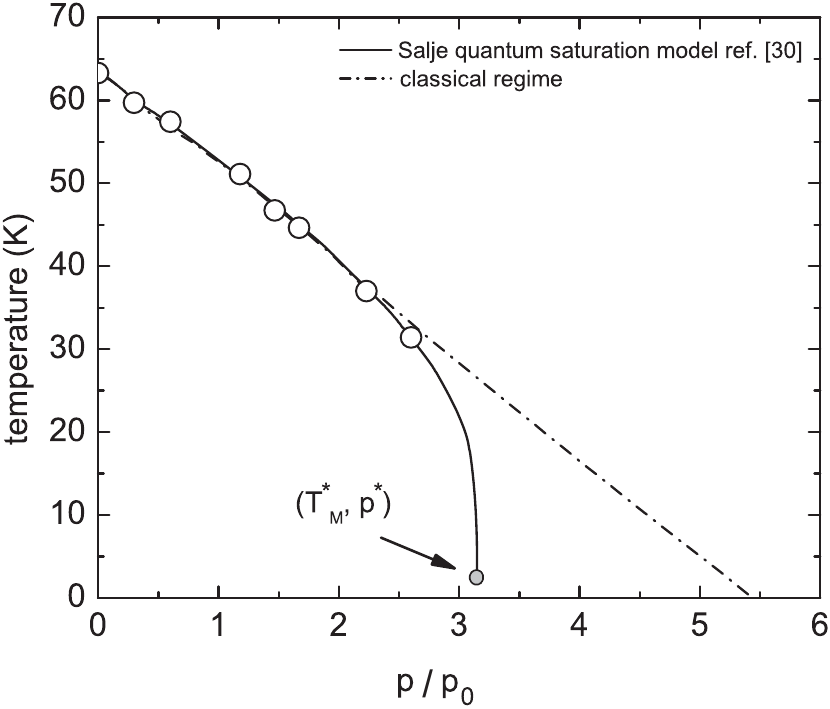}
   \caption{\label{Figure 6}
   Pressure dependence of the transition temperature up to pressures
   of nearly 3~GPa, together with a fit to Salje's model~\cite{Qsat}
   which accounts for quantum saturation effects close to zero absolute temperature.
   The difference between the linear (classical) and quantum-dominant behavior is illustrated in the inset.
   We find that quantum effects become important for temperatures lower than $\theta_s$~=~48.3$\pm$3.7~K,
   and a critical endpoint is predicted at $T_M^*$~=~2.7~K and
   $p^*$~=~3.1~GPa.
   }
\end{figure}

We use a modified GL free energy expression to describe the quantum
saturation in the order parameter expected at sufficiently low
temperature~\cite{Qsat}. This saturation effect reflects the
departure from classical behavior as the absolute temperature
approaches zero and is a direct consequence of the third law of
thermodynamics. Following Salje \emph{et al.}~\cite{Qsat}, we fit
the pressure dependence of the transition as $\theta_s/T_M(p) =
\coth^{-1} \bigl [ \coth(\theta_s/T_M(0)) - p/p_0 \bigr ]$. Here
$\theta_s$ is a phenomenological temperature below which quantum
mechanical effects are dominant. We find $p_0$~=~5.6$\pm$0.7~GPa and
$\theta_s$=48.3$\pm$3.7~K. The resulting phase diagram predicts a
critical endpoint at $T_M^*$~=~2.7~K and $p^*$~=~3.1~GPa.
Figure~\ref{Figure 6} depicts the experimental pressure dependence
of the transition temperature for pressures up to nearly 3~GPa,
together with the fit to Salje's classical regime. We find a
classical slope of -11.7$\pm$0.2~K GPa$^{-1}$. The difference
between the classical and quantum regimes is important for
temperatures below~$\theta_s$.

In conclusion, we report results of a detailed experimental study of
the martensitic transition in AuZn. While the temperature dependence
of the satellite peak intensity (proportional to the square of the
order parameter) show marked first-order behavior at 45~K for
Au$_{0.52}$Zn$_{0.48}$, we observed a continuous feature
well-described by a mean-field exponent ($\beta$=0.51) for AuZn at
64~K. This result is contrary to the established definition of a
martensitic transition. Providing supporting evidence to the
continuous nature of the phase transition in AuZn are the lack of
thermal hysteresis from electrical-resistivity measurements, the
$\lambda$-anomaly in the specific heat, and the disagreement of
pressure data with the Clausius-Clapeyron relation. Using Salje's
model to describe the quantum saturation effects close to zero
absolute temperature, we predict the presence of a critical endpoint
in the phase diagram located at $T_M^*$~=~2.7~K and $p^*$~=~3.1~GPa,
with a phenomenological temperature $\theta_s$=48.3$\pm$3.7~K for
the onset of quantum effects.

\begin{acknowledgments}
This work was performed under the auspices of the United States
Department of Energy and Department of Commerce and supported in
part by the Trustees of Boston College. Work at Brookhaven is
supported by the Office of Science, United States Department of
Energy under Contract No. DE-ACO2-98CH10886. The use of the Advanced
Photon Source was supported by the U.S. Department of Energy, Office
of Science, Office of Basic Energy Sciences, under Contract No.
DE-AC02-06CH11357.
\end{acknowledgments}

\end{document}